\documentclass[10pt, final, journal, letterpaper, twocolumn]{IEEEtran}

\usepackage{cite}
\usepackage{amsmath,amsthm,amssymb,amsfonts}
\usepackage{algorithm}
\usepackage{algorithmic}
\usepackage{graphicx}
\usepackage{color}
\newcommand{\bs}{\boldsymbol}
\usepackage[normalsize]{subfigure}
\usepackage{framed}

\DeclareMathOperator*{\argmax}{argmax}

\begin{document}

\title{Noise Is Useful: Exploiting Data Diversity for Edge Intelligence}
\author{Zhi Zeng, Yuan Liu,~\IEEEmembership{Senior Member,~IEEE}, Weijun Tang,~\IEEEmembership{Member,~IEEE}, and Fangjiong Chen,~\IEEEmembership{Member,~IEEE}
\thanks{
Z. Zeng, Y. Liu, and W. Tang are with the School of Electronic and Information Engineering, South China University of Technology, Guangzhou, 510641, P. R. China (email:  eezhizeng@mail.scut.edu.cn;  eeyliu@scut.edu.cn;  tangwj@scut.edu.cn). F. Chen is with the School of Electronic and Information Engineering, South China University of Technology, Guangzhou 510641, China, and also with the Key Laboratory of Marine Environmental Survey Technology and Application, Ministry of Natural Resources, Guangzhou 510300, China (e-mail: eefjchen@scut.edu.cn).
}
}

\maketitle

\vspace{-1.5cm}

\begin{abstract}
Edge intelligence requires to fast access distributed data samples generated by edge devices. The challenge is using limited radio resource to acquire massive data samples for training machine learning models at edge server. In this article, we propose a new communication-efficient edge intelligence scheme where the most useful data samples are selected to train the model. Here the usefulness or values of data samples is measured by data diversity which is defined as the difference between data samples. We derive a close-form expression of data diversity that combines data informativeness and channel quality. Then a joint data-and-channel diversity aware multiuser scheduling algorithm is proposed. We find that noise is useful for enhancing data diversity under some conditions.
\end{abstract}


\section{Introduction}

With the explosive increase of mobile devices and ubiquitous intelligent applications, massive data generated by edge devices materialize artificial intelligence (AI) or machine learning at network edge, known as \emph{edge AI} or \emph{edge intelligence}. However, many mobile devices, like internet-of-things (IoT) nodes, typically have small hardware-sizes and limited computational power. Thus the input data of edge devices are usually transmitted  via wireless links to an external computing system (i.e., edge server) for processing \cite{1,2,3}. As chips become more and more powerful, the computational power of edge server can be rapidly increased, wireless communication becomes a bottleneck to fast access the distributed data across edge devices. Moreover, the wireless transmission of high-dimensional training data from a large number of edge devices may congest the air interface due to limited radio resources. To overcome this challenge, it urges efficient wireless data transmission solutions for edge intelligence \cite{4,5,6,8}.

On one hand, the goal of communication is data rate maximization, in which channel is a ``bit-pipe" and data bits have equal value. However, in machine learning some data are more valued than the others. Thereby, to achieve communication-efficient edge intelligence, it is necessary that the edge server selects most valued or useful training data samples by limited radio resources. The idea of data selection comes from \emph{active learning} \cite{7}, where most \emph{informative} data samples are selected to be labeled (because manual labeling is costly), so as a model can be accurately trained by fewer labeled data samples. Based on this, the importance of data are differentiated in edge intelligence system  in \cite{9}, where the wrongly classified data and the local models trained by larger datasets are regarded to be more informative, and corresponding radio resource allocation schemes are designed. In \cite{10}, the data samples closer to the the decision boundary of support vector machine (SVM), i.e., the data with shorter distances to the decision boundary, are considered to be more informative.

On the other hand, noise is harmful in communication since it causes decoding error and thus makes communication unreliable. However,  reliability may not matter in machine learning. For example, when training neural networks, adding noise can help to avoid overfitting or being trapped in local solution, and improves training performance \cite{11,12}.

\begin{figure}[t]
\centering
\includegraphics[width=0.95\linewidth]{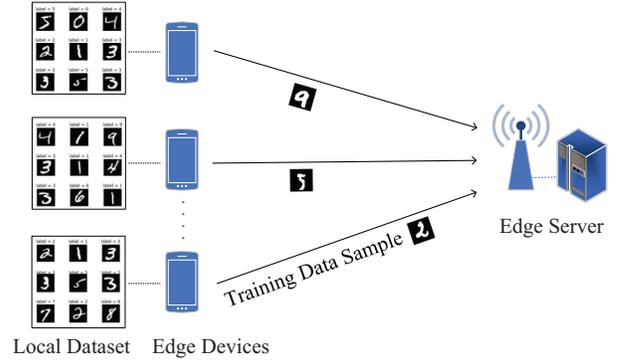}
\caption{System model of data selection in edge intelligence.}
\label{fig:wireless_model}
\end{figure}

In this article, we consider an edge intelligence system as shown in Fig. \ref{fig:wireless_model}, consisting one edge server and multiple edge devices. A certain machine learning model is trained at the edge server by using the data transmitted from the edge devices. The aim is to enhance the accuracy  and generalization of the model by using fewer radio resources. We propose a new data selection scheme by exploiting data diversity. That is, the edge server prefers to selecting the data samples that are most different from those that have been trained. We derive an explicit expression of the proposed data diversity metric, which interweaves the received signal-to-noise ratio (SNR) from communication and data distance from machine learning.  Different from the priori work \cite{9,10} that rely on model downloading at devices to evaluate data samples, in our scheme the devices only need to know the mean-value of the previously trained data samples and thus is model-free. We reveal that noise is useful under some conditions. Specifically, when the received SNR performance in the edge server is good, the added noise could enlarge data diversity and improve the performance of the trained  model.

The remainder of this article is organized as follows. Section \ref{sec:central} describes the system model of edge intelligence. Section \ref{sec:principle} presents the proposed scheme. Section \ref{sec:sim} provides experimental results and Section \ref{sec:con} finally concludes this article.

\section{System Model of Edge Intelligence}\label{sec:central}

We consider an edge intelligence system including an edge server and $K$ edge devices, where the edge devices transmit their individual labeled data samples to the edge server for training a machine learning model. The data sample transmission from the edge devices to the server is based on time-division manner and scheduled by the edge server.


Each device has a local dataset containing labeled training samples. Specifically, let $(\boldsymbol{x_k}, c_k)$ denote a labeled data sample of device $k$, with $\boldsymbol x_k$ representing the data sample and $c_k \in\{1,2,\cdots,C\}$ its corresponding label.
We consider a noisy data channel for high-rate data sample transmission and a label channel for corresponding label transmission. The latter is assumed to be noise-less for simplicity. This is reasonable since a label has a much smaller size than a data sample, e.g., a label is an integer of $0\sim9$ while a data sample is a vector of million coefficients. As time-division transmission is adopted, each slot is used to transmit a data sample of a scheduled device. We also assume that the data channel follows block-fading, i.e., the channel remains static within each slot but may vary from one slot to another. Due to the wireless fading and noise, the edge server receives biased training data samples sent from the edge devices. Therefore, if edge device $k$ is scheduled to transmit its data sample $\boldsymbol x_k$ at an arbitrary slot, the received signal at the edge server can be expressed as

\begin{equation}
\bs y_{k}=\sqrt{P}h_{k}\boldsymbol x_k+\boldsymbol z_k,
\label{eq:trans}
\end{equation}
where $P$ is the transmit power, $h_{k}$ is the channel gain from device $k$ to the edge server and $\boldsymbol z_k$ is the \emph{additive white Gaussian noise} (AWGN) vector following the  independent and identically distribution  (i.i.d.) $\mathcal{CN}(0,\sigma^{2})$. By multiplying $h_k^*$ to \eqref{eq:trans}, we can get:

\begin{align}
h_{k}^{*}\boldsymbol{y}_{k}&=\sqrt{P}h_{k}h_{k}^{*}\boldsymbol x_k+h_{k}^{*}\boldsymbol z_k \nonumber\\
&=\sqrt{P}|h_k|^2\boldsymbol x_k + h_{k}^{*}\boldsymbol z_k.
\label{eq:infer}
\end{align}

Considering analog transmission and maximum-likehood detection, the edge server decodes the received data sample as:

\begin{equation}\label{eq:rec_x}
\hat{\boldsymbol x}_k=\frac{1}{\sqrt{P}}\mathfrak{R}\left(\frac{h_{k}^{*}\boldsymbol{y}_k}{|h_{k}|^{2}}\right),
\end{equation}
where the real part of the received signal is extracted, since the training data samples are usually real-valued for machine learning. Thus, only the real part of the noise with $\sigma^2/2$ affects the data sample, and the received SNR for device $k$ is

\begin{equation}
\mathrm{SNR}_{k}=\frac{2P}{\sigma^{2}}|h_{k}|^{2}.
\label{eq:snr_cac}
\end{equation}

\section{Data-Diversity Aware Multiuser Scheduling}\label{sec:principle}

In this section, we derive a joint data-and-channel diversity policy for edge intelligence system. Since the radio resources are limited, to enable fast learning, it is crucial that the edge server schedules most useful data samples so as to ensure higher accuracy and fast convergence of machine learning model training. The policy design lies in a combination of wireless communication and machine learning, and it needs to take into account the both factors. 

\subsection{Diversity Metric}
The principle of active learning provides a relation between data diversity and model convergence: if highly disparate data are selectively added to the training set, better performance can be achieved with fewer data samples.
Take image classification as an example in machine learning, every pixel of an image belongs to the same attribute since it is represented by a gray value of size $0\sim255$. Thus we can calculate the Euclidean distance\footnote{Euclidean distance is a measure of the distance between two points in Euclidean space, with larger distance indicating more variation between two points. In this paper, we adopt the most popular Euclidean distance as an example to exhibit our data-diversity-aware scheduling scheme. Other measures, such as cosine similarity, Chebyshev distance, KL scatter and so on, can also be used to measure data distances depending on specific learning tasks, and our scheme is applicable for any measures of data distance.} between two data samples, and the value of this distance represents the difference between two data samples. If the difference between two data samples is larger, it means that information redundancy between the two data samples is smaller and thus more informativeness can be obtained for training machine learning models.

Based on above argument, the data diversity is measured by the Euclidean distance between data samples. Given any data samples $\boldsymbol{x}_{1}$ and $\boldsymbol{x}_{2}$, their distance can be readily computed by $d(\boldsymbol x_1, \boldsymbol x_2)={||\boldsymbol{x}_{2}-\boldsymbol{x}_{1}||}_{2}$.
Then the distance based data-diversity measure is defined as

\begin{equation}
\\ d^2(\boldsymbol{x}_1,\boldsymbol{x}_2)=\left|\left|\boldsymbol{x}_{2}-\boldsymbol{x}_{1}\right|\right|_2^2.
\label{variety}
\end{equation}

However, the measure of data diversity in active learning is for noiseless data and cannot be used directly in edge intelligence where the received data samples at the edge server are corrupted by wireless fading and noise.

Therefore, the idea of our scheme is described as follows:  For transmitting a particular data sample, the transmitter (device) does not know the specific noise experienced by the transmitted data samples. As the data selection metric is computed at each transmitter-side, we consider the statistical properties of noise so as to predict how a data sample to be scheduled is affected by noise. To this end, we take expectation over the noise of the received data sample, since only the noise is random and uncertain for the received signal \eqref{eq:trans}.  Note that in our experiments the noise is randomly added to data samples.  Specifically, denote $\overline{\boldsymbol x}_0$ as the central point (or the mean) of the training data samples that are received at the edge server in previous slots, then the distance from any data sample ${{\boldsymbol{\hat{x}}_{k}}}$ to $\overline{\boldsymbol x}_0$ is given by:

\begin{align}\label{eq:data_diversity}
d(\hat{\boldsymbol{x}}_{k}, \overline{\boldsymbol x}_0)&={{||\hat{\boldsymbol{x}}_{k}-\overline{\boldsymbol x}_0||}_{2}} \nonumber\\
&={{\left|\left|(\frac{h_{k}^{*}\boldsymbol{y}_k}{\sqrt{P}|h_{k}|^{2}})-\overline{\boldsymbol x}_0\right|\right|}_{2}} \nonumber\\
&={{\left|\left|(\frac{\sqrt{P}|h_{k}|^{2}\boldsymbol{x}_k+h_{k}^{*}\boldsymbol{z}_k}{\sqrt{P}|h_{k}|^{2}})-\overline{\boldsymbol x}_0\right|\right|}_{2}} \nonumber\\
&={{\left|\left|(\boldsymbol{x}_{k}+{\frac{h_k^* \boldsymbol z_k}{\sqrt{P} |h_k|^2}})-\overline{\boldsymbol x}_0\right|\right|}_{2}},
\end{align}
where $\hat{\boldsymbol x}_k$ is the decoded data sample defined in \eqref{eq:rec_x}. By taking the expectation operation for \eqref{eq:data_diversity} over the random variable $\boldsymbol z_k$, we have the closed-form expression of the diversity as:

\begin{framed}
\noindent
(\textbf{Joint Data-and-Channel Diversity Measure}):

\begin{equation}\label{eq:d_cac}
\mathbf{E}_{\boldsymbol z_k}\left[d^{2}(\hat{\boldsymbol{x}}_{k}, \overline{\boldsymbol x}_0)\right]=d^{2}(\boldsymbol x_k, \overline{\boldsymbol x}_0)+\frac{1}{{\mathrm{SNR}_{k}}}.
\end{equation}

\end{framed}

\begin{figure}[t]
\centering
\includegraphics[width=0.94\linewidth]{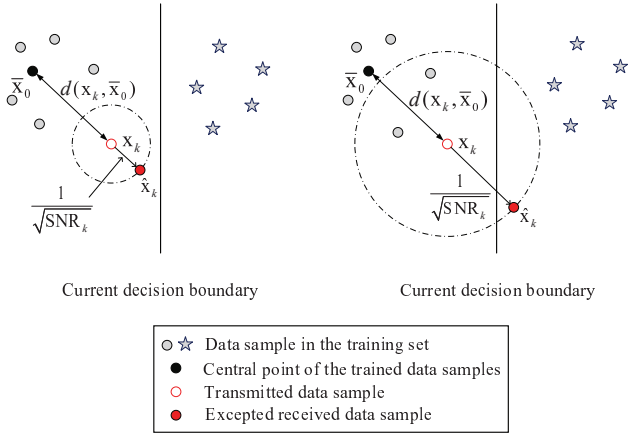}
\caption{An example of the effect of noise.}
\label{fig:SVM_model}
\end{figure}

Here \eqref{eq:d_cac} shows that channel decay and noise can affect the data diversity, which are reflected together by the term of SNR and can increase the data diversity. The effect of noise on the data diversity can be further illustrated in Fig. \ref{fig:SVM_model}, where a SVM classifier is adopted as an example. It can be observed that the noise provides randomness to data samples and enlarges data diversity, which is benefit to the generalization ability of the trained model. But if the noise is too strong, the transmitted data sample may be far away from its noise-less position and becomes a misclassified sample. Our result is also of great practical significance. Enlarging data-diversity is one of the promising methods to improve the performance of a learning model. For example, in computer vision, the original training samples are often manually modified via rotation, flipping and many other transformations to enlarge the training dataset. In this article, we exploit the inherent received noise in wireless communications to enlarge the dataset and data-diversity.


\begin{algorithm}[!t]
\caption{Data-Diversity Aware Multiuser Scheduling}
\begin{algorithmic}[1]
\STATE \textbf{initialize} The received data sample set $\mathcal D$ and $\overline{\boldsymbol x}_0$.
\REPEAT

\STATE \emph{(Central Point Broadcasting):} The server broadcasts $\overline{\boldsymbol x}_0$ to all devices;
\STATE \emph{(Diversity Measure):} Each device $k$ calculates and then uploads the measure $I_k$ to the server;
\STATE \emph{(Transmission Scheduling):} The server selects the device $k^*$ with the maximum $I_k$ for transmitting data $\boldsymbol x^{*}_{k^*}$;
\STATE \emph{(Updating):} $\mathcal D \leftarrow \mathcal D \cup \hat{\boldsymbol x}^{*}_{k^*}$;

$\quad\quad\quad\quad\quad \overline{\boldsymbol x}_0 \leftarrow \frac{1}{|\mathcal D|}\sum_{{\boldsymbol x}\in\mathcal D}{\boldsymbol x}$;

$\quad\quad\quad\quad\quad$Train a new model by using dataset $\mathcal D$;

 \UNTIL{Model converges or transmission budget exhausts. }
\end{algorithmic}
\end{algorithm}

\subsection{Multiuser Scheduling}

The diversity expression \eqref{eq:d_cac} combines both communication and machine learning to reveal the usefulness of a data sample for machine learning. Thus, at each transmission, each device $k$ prefers to selecting one data sample from its local dataset to achieve the maximum diversity in \eqref{eq:d_cac}, in which the best data sample of each device $k$ is denoted as $\boldsymbol x_k^*$. Then, each device $k$ uploads its diversity measure $I_k$ to the edge server:

\begin{equation}\label{eq:I_k}
I_{k}=\frac{1}{\mathrm{SNR}_{k}}+ d^2(\boldsymbol x_k^*, \overline{\boldsymbol x}_0).
\end{equation}

After receiving the measures $I_k$'s from all the devices, the edge server has the following scheduling policy.

\begin{framed}
\noindent
(\textbf{Joint Data-and-Channel Diversity Scheduling}): At each transmission time, the edge server schedules device $k^{*}$ to upload a data sample if

\begin{equation}\label{schedule1}
k^{*}=\argmax_{k}\left\{\frac{1}{\mathrm{SNR}_{k}}+ d^2(\boldsymbol x_k^*, \overline{\boldsymbol x}_0)\right\}.
\end{equation}

\end{framed}

Finally, we formally describe the whole scheme in Algorithm 1. At the first step, the server broadcasts the central point $\overline{\boldsymbol x}_0$ of the set of the trained data samples $\mathcal D$. Then, each device $k$ calculates its diversity measure $I_k$ by selecting one best data sample from its local dataset, according to \eqref{eq:I_k}.   The server schedules one best device $k^*$ that has largest $I_k$ and updates the central point $\overline{\boldsymbol x}_0$. The above steps are iterated until the model converges or transmission budget exhausts.

\section{Experimental Results}\label{sec:sim}

In this section, we evaluate the proposed scheme via experiments.

\subsection{Experimental Settings}
We adopt SVM as an example of the machine learning model for the purpose of illustration. Note that the proposed scheme is applicable for any algorithms in machine learning. In order to reduce the consumption of wireless communication resources, binary soft-margin SVM model is used. Initially, we build an original classifier model with some initial data samples that are already on the server before collecting wireless data samples distributed across the edge devices. As more and more training data samples are uploaded to the edge server by the edge devices, the classifier model is gradually corrected, and its ability to classify correctly continues to improve.

We consider $20$ edge devices unless specified otherwise.
The wireless fading $h_k$'s are assumed to be Rayleigh fading. We use the well-known MNIST dataset of handwritten digits to train the SVM classifier, which consists of two parts: a training set containing $60,000$ samples and a test set containing $10,000$ samples, and each set comprises data and labels. Each data in the MNIST data set is a gray image of $28\times28$ pixels, which means that the dimension of a data is $784$, corresponding to $784$ columns in the data matrix, and each row is a gray image. The content of these data are handwritten numbers 0$\sim$9, and these $10$ categories correspond to the $10$ columns of the label matrix respectively, while each row represents the corresponding image located in the same row of the data matrix. In every row, only one column that the category belongs to is marked as $1$, and the others are marked as $0$. In order to highlight the results of the experiment, we select two categories, $3$ and $5$, from the entire dataset, which included a total of $11,552$ training samples and $1,902$ test samples. The initial model is constructed from a small number of samples stored in the edge server. The remaining training data are randomly and uniformly distributed on the edge devices to build the local datasets.

For comparison, we also investigates three benchmarks. The first benchmark only considers data diversity and ignores the communication factor, i.e., the device scheduling policy is $k^*=\arg\max_{k} d^2(\boldsymbol x_k^*, \overline{\boldsymbol x}_0)$. The second benchmark is based on multiuser diversity, where each transmission schedules a device with the maximum channel gain, i.e., $k^*=\arg\max_{k} |h_k|^2$. The last one is random scheduling, i.e., at each slot, the server randomly schedules a device that randomly selects one data sample.

\subsection{Learning Performance}

\begin{figure}
\centering
\includegraphics[width=9.5cm]{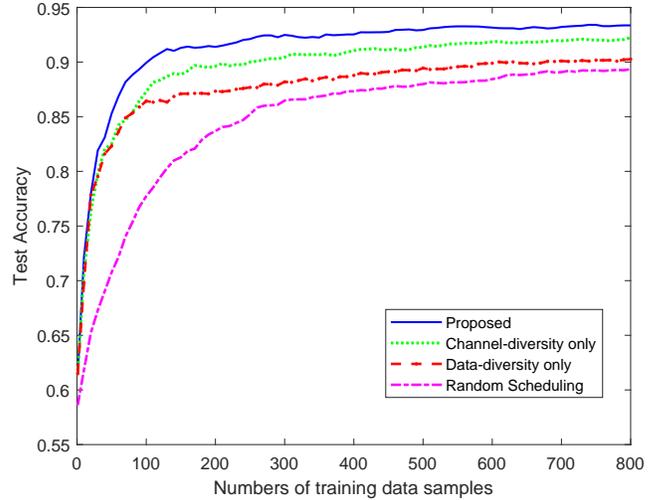}
\caption{Test accuracy versus transmission budget.}
\label{diff_scheme}
\end{figure}

\begin{figure}
\centering
\includegraphics[width=9.5cm]{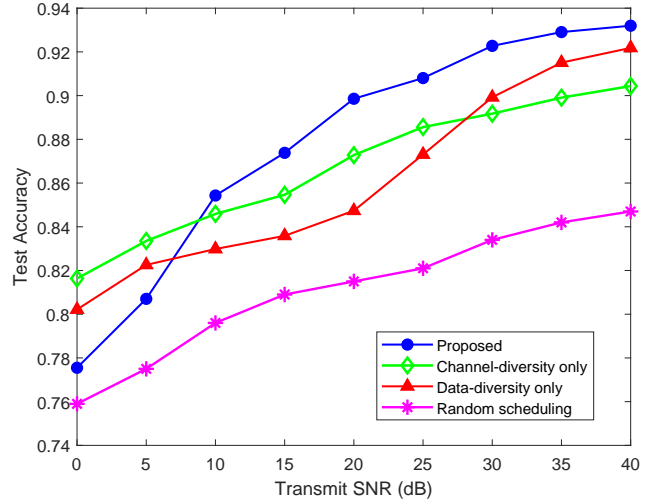}
\caption{Test accuracy versus the average transmit SNR.}
\label{diff_SNR}
\end{figure}

\begin{figure}
\centering
\includegraphics[width=9.5cm]{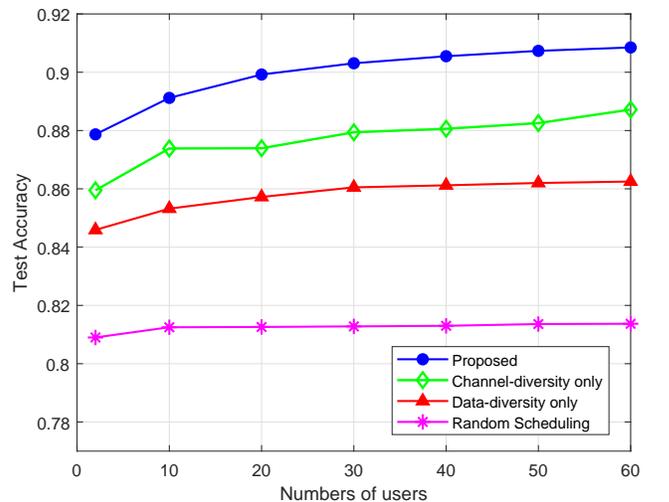}
\caption{Test accuracy versus the numbers of users.}
\label{fig:diff_user}
\end{figure}

\begin{figure}
\centering
\includegraphics[width=9.5cm]{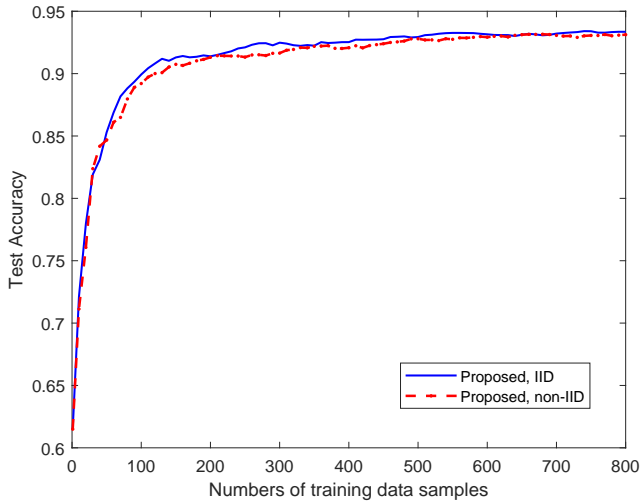}
\caption{Test accuracy versus the transmission budget.}
\label{fig:noniid}
\end{figure}

\subsubsection{Convergence Rate}
In Fig. \ref{diff_scheme}, we investigate the learning performance of the three schemes, where the transmit SNR $P/\sigma^2$ is set to be $20$ dB for all devices. The total transmission budget is set as $800$.  We can see that the proposed scheme is significantly better than the other three benchmarks. Though the three benchmark schemes are able to converge, they require much more transmission resources than the proposed scheme, e.g., the first two benchmark schemes consume $2$ and $6$ times of resources than the proposed scheme, respectively, when the test accuracy is set as $0.9$ in this example. In contrast, the random scheduling scheme can not achieve up to $0.9$ test accuracy. This confirms that the proposed scheme exploiting both data diversity and channel quality results in rapid convergence of the model. It validates the effectiveness of the proposed solution for fast learning.

\subsubsection{Transmit SNR}
To check the robustness to channel conditions, the all schemes are tested at different transmit SNR and the results are shown in Fig. \ref{diff_SNR}, where the transmission budget is fixed as $200$. It can be observed that the test accuracy of the proposed scheme is lower than three benchmark schemes at low SNR. However, as the transmit SNR increases, the test accuracy gradually increases and eventually exceeds the other schemes. This is consistent to our above analysis that large noise (or low SNR) will make the received data samples to be a wide range of deviations from the expected data samples so that the model is trained by the wrong data samples. But, small noise allows the server to obtain data samples with larger diversity and thus accelerate model convergence. Moreover, as the transmit SNR improves, the test accuracy of the benchmark scheme with data-diversity only exceeds the channel-aware scheduling scheme, eventually approaching the proposed scheme.  Moreover, the random scheduling scheme has the worst performance regardless of the SNR regimes.

\subsubsection{Multiuser Diversity}
We also analyze the user/channel diversity gain by plotting the test accuracy for different numbers of devices, as shown in Fig. \ref{fig:diff_user}, where the transmission budget is fixed as $200$ and the transmit SNR is fixed as $20$ dB. The performance of proposed scheme outperforms the three benchmarks. When the number of users is small, there are less training data samples in the edge devices that the server can select. As more devices enter the system,  both data and channel diversities that are available to the system increases.  It is noted that the performance of the random scheduling scheme can not explore the data and channel diversities since both data and user are randomly selected.

\subsubsection{Non-IID Data Distribution}
Data imbalance or non-IID data distribution problem is usually encountered in machine learning. To verify the performance of the proposed data diversity aware scheme in non-IID case, we conduct experiments and the results are show in Fig. \ref{fig:noniid}, where we can observe that the performance of the non-IID case is slightly worse than that of the IID case. This also shows the robustness of our proposed data-diversity aware scheme. This is because the proposed scheme aims at selecting the most different data sample (compared with the average point of the trained data samples at the server) instead of a certain class of data in each iteration, the uneven data distribution on devices does not fundamentally affect the selection. Therefore, our proposed scheme is robust under the non-IID data distribution.

\section{Concluding Remarks}\label{sec:con}
In this article, we proposed a new scheduling scheme that exploits data diversity besides communication reliability. The proposed scheme selects the most useful data samples measured by data diversity for model training so as to accelerate the training process. The proposed scheme can be extended to more sophisticated scenarios of wireless communication.

\bibliographystyle{IEEEtran}
\bibliography{IEEEabrv,reference_paper}

\end{document}